\documentclass{jltp}

\usepackage{graphicx}

\title{Horizons and Ergoregions in Superfluids}

\author{G.E. Volovik\address{Low Temperature Laboratory, 
Helsinki University of Technology\\
P.O.Box 2200, FIN-02015 HUT, Finland\\
Landau Institute for Theoretical Physics, 
Kosygina 2, 119334  Moscow, Russia }}

\runninghead{G.E. Volovik}{Horizons and Ergoregions in Superfluids}
       
\begin{document}

\maketitle

\begin{abstract}
Ripplons -- gravity-capillary waves on the free surface of
a liquid or at the interfaces between two superfluids -- are the most
favourable excitations for simulation of the general-relativistic effects
related to horizons and ergoregions.  The white-hole horizon for the
``relativistic'' ripplons at the surface of the shallow liquid is easily
simulated using the kitchen-bath hydraulic jump. The same white-hole
horizon is observed in quantum liquid -- superfluid $^4$He. The
ergoregion for the ``non-relativistic'' ripplons is generated in the
experiments with two sliding $^3$He superfluids. The common property
experienced by all these  ripplons is the Miles instability inside the
ergoregion or horizon. Because of the universality of the Miles
instability, one may expect that it could take place inside the horizon
of the astrophysical black holes, if there is a preferred reference frame
which comes from the trans-Planckian physics. If this is the case, 
the black hole would evapotate much faster than due to the Hawking
radiation. Hawking radiation from the artificial black hole in terms of
the quantum tunneling of phonons and ripplons is also discussed.

\end{abstract}


\section{INTRODUCTION}

Black hole is the region of  spacetime in which gravity is so strong that
not even light can escape from this region. There are many astrophysical
black hole candidates, but the unambiguous
observational evidence for the existence of black holes has not yet been
established (see, e.g. review papers \cite{AstronomicalBH}). The defining
property of the black hole -- the event horizon -- is still missing in the
experiment, though there are indications that some of the candidates might
have event horizons.  At the moment we are not able  to study
experimentally the exotic phenomena related to the event horizon, for
example, how the presence of the horizon modifies the properties of the
vacuum of relativistic quantum fields. That is why the modeling of the
event horizon in condensed matter systems may provide us with some ideas
useful for study of the astronomical black holes, their formation and
decay. The close analogy between the relativistic quantum vacuum and the
ground state of quantum condensed matter appears to be instrumental.  In
this respect the quantum liquids, such as superfluids, are the best
candidates for simulations:  the horizon can be relatively easy
constructed in the moving liquid, and the properties of the quantum
vacuum in the presence of the horizon can be studied.

 Two-fluid  hydrodynamics of superfluid  liquids describes the coherent
motion of the superfluid `quantum vacuum' and  dynamics of excitations
above the ground state -- quasiparticles -- which form the  normal
component of the liquid.  At low energy some quasiparticles mimic 
relativistic particles propagating in the effective  curved space-time
produced by the distortions of the superfluid vacuum and by its 
superfluid flow.  This is an example of  the quantum gravity provided by 
quantum liquids.  Starting from the trans-Planckian physics --  quantum
many body system of interacting atoms -- one obtains in the low energy
corner the emergent  quantum fields of relativistic quasiparticles 
living in the background of the emergent classical gravity field. 

The effective gravity  in quantum liquids is dynamical,  but as distinct
from the Einstein theory of gravity, the metric field typically obeys the
hydrodynamic equations rather than Einstein equations. There is an
exclusion from this rule:  in superfluid $^3$He-A, the effective quantum
electrodynamics of ``relativistic''  fermions and ``photons''  emerges at
low energy. These quantum fields propagate in effective space-time  whose
metric field obeys  equations remotely resembling Einstein equations
\cite{VolovikBook}. All this occurs because superfluid $^3$He-A belongs
to the special universality class of Fermi systems.

However, even if the effective gravity in liquids is  distinct from the
Einstein gravity, it can be useful for simulations of many phenomena
related to Einstein gravity, including the physics of black holes.   For
example, the effect of Hawking radiation is solely determined by the
behavior of quantum fields in the vicinity of the event horizon, and  it
does not matter  for which excitations (photons or sound waves) the
horizon is constructed and from which equations (Einstein or
hydrodynamic) it is obtained.  

The acoustic analog of a black hole -- the
horizon for sound waves -- has been suggested by Unruh in 1981 
\cite{Unruh1981}. The event horizon for sound waves emerges if the flow
velocity of the liquid exceeds
the local speed of sound $c$ in some region of liquid: sound waves (or
quasiparticles in superfluids -- phonons) cannot escape from this region.
The propagation of sound in the presence of the 
acoustic horizon is described by the so called acoustic metric, which is
similar to the metric describing the space-time in general relativity in
the presence of the black hole. This is discussed in Sec.
\ref{SecGravityphonons} together with the analog of the Hawking radiation
which is described in terms of quantum tunnelling between the classical
trajectories
\cite{SrinivasanPadmanabhan,VolovikTunneling,ParikhWilczek}. Since 1981 there appeared
many suggestions to simulate the black hole and white hole horizons for
various excitations in various laboratory systems (see review paper 
\cite{VisserReview} and references therein).
 
In Sec. \ref{SecGravityRipplons}  we discuss the most perspective  analog
--  the effective 2+1 space-time  emerging for the surface waves 
(ripplons) in the shallow water limit. The effective horizon for ripplons
has been first suggested by Schutzhold and Unruh \cite{SchutzholdUnruh}. 
This analog demonstrates the mechanism of the decay of the black hole
which is alternative to the Hawking radiation -- the instability of the
quantum vacuum behind the horizon (Sec.
\ref{Instability}). This mechanism  is applicable both to the
relativistic systems with the horizon and non-relativistic systems with
the ergoregion, and it has been experimentally observed for the
non-relativistic ripplons propagating at the interface between two
superfluids  
 \cite{Kelvin-HelmholtzInstabilitySuperfluids,ROPrevirew,Volovik}.   
Similar instability of the vacuum inside the astronomical black hole is
possible.  

In Sec. \ref{WH} we discuss  experiments with the hydraulic jump    in
superfluid  $^4$He \cite{Pettersen}. This circular hydraulic jump
simulates the 2+1 dimensional white hole for the  relativistic ripplons 
and the instability inside the horizon \cite{HJumpWH}.   

In Sec. \ref{ConclusionSec} some perspectives are discussed.

\section{GRAVITY FOR PHONONS}
\label{SecGravityphonons}

\subsection{Effective Metric}
\label{SecEffectiveMetricPhonon}

In the frame moving with velocity ${\bf v}_{\rm s}$ of the  superfluid
vacuum -- the comoving frame -- the  spectrum of sound waves is
`relativistic': $\omega^{\rm com}({\bf k})=\pm ck$, where $c$ is the
speed of sound. In the laboratory frame, the  spectrum is Doppler shifted:
\begin{equation}
\omega({\bf k},{\bf r})=
\pm ck+{\bf k}\cdot{\bf v}_{\rm s}({\bf r})~,
\label{GalileanTransform}
\end{equation}
or
\begin{equation}
\left(\omega-{\bf
k}\cdot{\bf v}_{\rm s}\right)^2
=c^2 k^2~.
\label{EbasicPhonons}
\end{equation}
This can be written  in the general Lorentzian form  
\begin{equation}
g^{\mu\nu}k_\mu k_\nu = 0~,
\label{Contra1}
\end{equation}
 where $k_0=-\omega $, and
the contravariant components of the metric are

\begin{equation}
g^{00}=-1~,~    g^{0i}({\bf r})=  -v_{\rm s}^i({\bf r})~,~g^{ij}({\bf
r})= c^2 \delta^{ij} -v_{\rm s}^i({\bf r}) v_{\rm s}^j({\bf r})
~.
\label{Contra2}
\end{equation}
The inverse matrix $g_{\mu\nu}$ determines the trajectories of phonons in
the moving liquid. This covariant metric describes the effective
(3+1)-dimensional space-time in which phonons move along the geodesic
lines $ds^2=0$:
\begin{equation}\label{Covariant}
ds^2=g_{\mu \nu}dx^\mu dx^\nu= -dt^2   +
c^{-2}  (d{\bf r}-{\bf v}_{\rm s} ({\bf r})dt)^2 ~.
\end{equation}

 \subsection{Black and White Holes}
\label{SecAcousticHorizon}

The most instructive for us is the spherically symmetric flow when the
superfluid velocity  is radial:  ${\bf v}_{\rm s}({\bf r})= \hat {\bf r}
v(r)$:  
\begin{equation}\label{PainlevGullstrand}
ds^2= -(1-v^2(r)/c^2)dt^2 -  2c^{-2}v(r)drdt   +
c^{-2}  (d{\bf r})^2 ~.
\end{equation}
The coordinate transformation
\begin{equation} 
 \tilde t=t + \int^r dr{  v(r) \over c^2 -v^2(r)}~.
\label{ForbiddenCoordinateTransformation}
\end{equation}
leads to the more familiar metric
\begin{equation} 
ds^2= -\left(1-{v^2(r)\over c^2}\right)d\tilde t^2  + {dr^2\over
c^2}  {1\over 1-{v^2(r)\over c^2} } + {1\over c^2} r^2 d\Omega^2 ~.
\label{SchwarzschildLineElement}
\end{equation}

In the particular case when the  velocity field has  the form
$v^2(r)=2GM/r$, the equation (\ref{SchwarzschildLineElement}) reproduces 
the Schwarzschild line element in Einstein gravity, where $M$ is the mass
of the gravitating body, and $G$ is the Newton constant. The
corresponding equation (\ref{PainlevGullstrand}) with $v^2(r)=2GM/r$ is
known as the Painlev\'e--Gullstrand line element which describes the
gravity field of the same body but in different coordinate system -- the
coordinate system adapted to the free falling particle \cite{Painleve}.

 In moving superfluids, the  Painlev\'e--Gullstrand type metric
(\ref{Covariant}) with nonzero off-diagonal element $g_{0i}$ is naturally
generated by flow since $g_{0i}\propto v_{{\rm s}i}$. The notion of the
black and white event horizons has also a very simple meaning in liquids.
For example, let us assume that we are able to construct  the spherically
symmetric flow  with $v(r)<0$ such that within some radius $r<r_h$  one
has  $v^2(r)>c^2$. Then phonons cannot escape from this region since they
are dragged by the superfluid vacuum to the center with the speed faster
than their velocity $c$. Thus the surface $r=r_h$ at which $v(r)=-c$
would serve as the horizon of the black hole. 

On the contrary, if the liquid flows outward, i.e. for $v(r)>0$,
all the
phonons are necessarily  dragged away from the  region  $r<r_h$. In this
case, the surface $r=r_h$ at which $v(r)=c$  would serve as the horizon
of the white hole. 
 
Though in the Einstein gravity the Schwarzschild metric and the
Painlev\'e--Gullstrand metric describe the same gravitational field, the 
latter is more preferable  especially  when the physics of the event
horizon is concerned.  The Schwarzschild metric contains the singularity
at the horizon:  the  $g_{rr}$ element of the metric  is infinite at
$v^2(r)=c^2$. This is the so-called coordinate singularity --  the
non-physical singularity which can be removed by coordinate
transformation to the Painlev\'e--Gullstrand metric. The latter  is
smooth across the horizon,  which allows us to study the behavior of the
quantum fields and quantum vacuum both outside and inside the horizon.
This is the reason why the Painlev\'e--Gullstrand metric which provides 
us with the simple ``river'' model of black hole  popular now in physics
of the event horizon (see Ref. \cite{RiverModel}   and references
therein).

 \subsection{Hawking Radiation as Tunneling}
\label{HawkingRradiationPainleveSec}
 
As distinct form the Einstein equations which admit the  spherically
symmetric solutions with the horizon,  the equations of hydrodynamics do
not support the spherically symmetric acoustic horizon. Acoustic horizon
occurs in a different geometry: in the so-called Laval nozzle, where the
horizon takes place at the narrowest cross section of the tube.
Fortunately, the shape of the surface of the horizon is not important for
the discussion of the most interesting quantum properties of the quantum
fields in the  horizon --  the Hawking radiation. This effect can be
illustrated using the simplest case of the one-dimensional flow with
varying velocity $v(x)$ and varying speed of sound $c(x)$. The effective
acoustic metric experienced by phonons has the form
\begin{equation}\label{Covariant1D}
ds^2=g_{\mu \nu}dx^\mu dx^\nu= -dt^2   +
{1\over c^2(x)}  (dx-v(x)dt)^2 ~,
\end{equation}
and the horizon occurs at the point $x=x_0$ where $v(x_0)=c(x_0)$,  or
$v(x_0)=-c(x_0)$. The effective gravitational field at the horizon  is
given by 
\begin{equation}\label{GravityHorizon}
 g_x=(1/2)\nabla_x(c^2-v^2)|_{x_0} 
~.
\end{equation}

Let us choose the flow  with $v(x)>0$. To simulate the black hole horizon,  
we must take $v(x)<c(x)$ at $x<x_0$ (exterior region) and $v(x)>c(x)$ at
$x>x_0$ (region behind the horizon).  In the local reference frame
comoving with liquid, the energy spectrum of phonons is everywhere
positive $\hbar \omega^{\rm com}(k)=\hbar  ck >0$, and thus at $T=0$
phonons are not excited. This vacuum state is, however, not in
equilibrium since in the local comoving frame the whole velocity field is
time dependent and thus the quasiparticle energy and the vacuum as a
whole is not well defined. The energy is well defined in the laboratory
frame where the velocity field is time independent. But in this frame
some modes acquire negative energy behind the horizon, $\hbar \omega({\bf
k})=\hbar  ck  +\hbar {\bf k}\cdot {\bf v}_{\rm s}<0$. All this indicates
that the initial comoving quantum vacuum is unstable in the presence of
the black hole.  Hawking  radiation provides the mechanism for the decay
of the black hole unless some other, more violent, process of vacuum
instability intervenes. 

\begin{figure}[t]
\centerline{\includegraphics[width=80.mm]{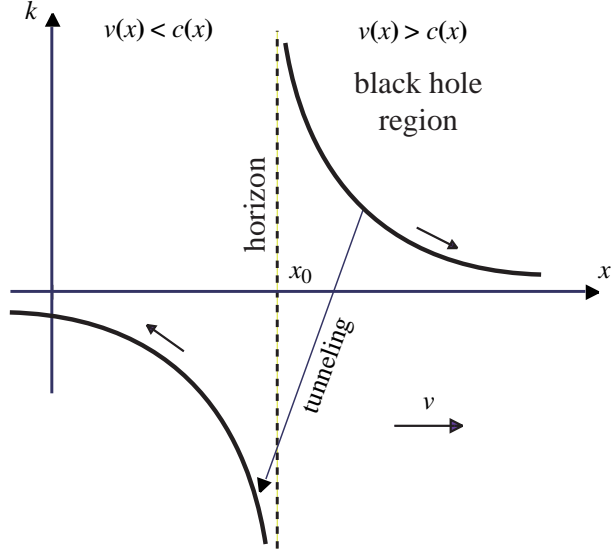}}
%
\caption{Two relevant rajectories of massless `relativistic'
quasiparticles -- phonons -- travelling from the horizon of acoustic
black hole. Arrows show the direction of motion. Quasiparticles have
diverging energy in the vicinity of the horizon (infinite blue shift)
which must be cut off by Planckian (Debye) energy scale. The tunneling
between the two trajectories describes the process of emission of phonons
from the horizon which is analogous to the Hawking radiation from the
black hole.} 
\label{TunnelingTrajJETPLFig}
 \end{figure}

In the semiclassical description, the Hawking radiation can be considered
as the quantum tunneling between classical trajectories
\cite{SrinivasanPadmanabhan,VolovikTunneling,ParikhWilczek} (the same analysis applied
for the Unruh effect in superfluids see in Ref. \cite{VolovikExotic}).  We consider the
positive frequency  modes,
$\omega=\omega(k)>0$,  as viewed in the laboratory frame. 
Classical trajectories $k(x)$ of phonons may be found from
Eq.(\ref{GalileanTransform}):
$\omega=\pm c(x)|k(x)|+ v(x)k(x)$. There are two relevant  trajectories
in Fig. \ref{TunnelingTrajJETPLFig}:
\begin{equation} 
 k(x)=\frac {\omega}{v(x)- c(x)} ~.
\label{Trajectories}
\end{equation}
 The trajectory with $k(x)<0$ starts at the horizon and  propagates into
the exterior region. In the comoving frame the phonon frequency is
positive, $\omega^{\rm com}= c(x)|k(x)|$. Formally the trajectory starts
with infinite frequency, $\omega^{\rm com}(x_0)= +\infty$, but
practically there is the cut-off which in quantum liquids is given by the
Debye frequency, and in the quantum  vacuum -- by the Planck energy
scale. The trajectory with $k(x)>0$ also starts near horizon but
propagates into the region behind the horizon.  In the comoving frame the
phonon frequency is negative, $\omega^{\rm com}(k)=- c(x)|k(x)|$, and
approaches $-\infty$ at the horizon.

 Quantum mechanics allows the phonon to tunnel between  the two
trajectories. The tunneling exponent is determined by the conventional
semiclassical action 
\begin{equation} 
S= {\rm Im}\int  \hbar k (x)dx=  {\rm Im}\int \frac{
\hbar\omega}{v(x)-c(x)}dx~,
\label{TunnelingExponent}
\end{equation}
The momentum $k(x)$ on the classical trajectory has a pole at the horizon
\begin{equation} 
 k(x)\approx\frac {\omega c(x_0)}{g_x(x-x_0)} ~,
\label{Pole}
\end{equation}
 Shifting 
the contour of integration into the upper half-plane of the complex
variable $x$, one obtains for the
tunneling action
\begin{equation}
S =\frac{\pi  \hbar\omega c(x_0)}{g_x}  ~~.
\label{HawkingTvelocity2}
\end{equation}
The tunneling between the trajectories describes the process of  creation
of the pair of phonons, one of which is radiated into the exterior
region. The probability of tunneling  demonstrates that  the radiation
from the (acoustic) black hole is thermal with the Hawking temperature
$T_{\rm H}$ being proportional to the (effective) gravity at the horizon:
 \begin{equation}
P\propto\exp\left(-{2S\over \hbar}\right)=
\exp\left(- { \hbar \omega\over T_{\rm H} }\right)~~,~~T_{\rm
H}=\frac{\hbar g_x}{2\pi c} ~.
 \label{HawkingTvelocityProbability}
\end{equation}
 This semiclassical approach is only valid  when the tunneling action is
large compared to Planck constant, i.e. when
$ \hbar \omega\gg T_{\rm H}$.

The Hawking radiation is completely determined  by the (effective) metric
in the vicinity of the horizon. It does not depend on the motion
equations  for the gravity field: the metric can obey Einstein,
hydrodynamic or other equations which allow for the existence of an event
horizon. The Eq.(\ref{HawkingTvelocityProbability}) is applicable for any
relativistic excitations radiated from the horizon, including
`relativistic' fermionic quasiparticles living in superfluid $^3$He-A
\cite{JacobsonVolovik,VolovikTunneling}.

Note that in the above derivation of the Hawking temperature we actually
did not use the full quantum mechanics. Only the wave mechanics was
exploited, which is also applicable to the classical waves propagating in
the background classical metric.
Eq.(\ref{HawkingTvelocityProbability}) can be rewritten without using the
Planck constant $\hbar$ (see Ref. \cite{NouriZonoz}):
 \begin{equation}
P\propto\\exp\left(-  \frac{\omega} {\omega_{\rm H}}\right)~~,
 ~~\omega_{\rm H}=\frac{ g_x}{2\pi c} ~.
 \label{ClassicalHawkingProbability}
\end{equation}
 The above equation describes the spectrum of classical  waves radiated 
from the horizon.  In the full quantum mechanics the radiation is caused 
by quantum fluctuations in the vicinity of the horizon.  But in the wave
mechanics the radiation can be generated by some external noise
introduced in the vicinity of the horizon or by thermodynamic
fluctuations.  However, these initial perturbations near the horizon are
only responsible for the prefactor in
Eq.(\ref{ClassicalHawkingProbability}), which was missing in our
derivation, but they do not influence the value of the Hawking frequency
$\omega_{\rm H}$.  That is why the Hawking frequency can be measured in
classical experiments.
 
\subsection{Does Hawking Radiation Exist?}
\label{HawkingRradiationExistenceSec}

Let us mention that in the literature there  is a controversy concerning the Hawking
radiation. There is even a point of view that a Schwarzschild black hole formed during
a collapse process does not radiate (see e.g. Ref. \cite{Belinski}). The other authors
(see e.g. Ref. \cite{Akhmedov}) suggest  that the  Hawking radiation may exist but the
Hawking temperature is twice larger than that originally calculated by Hawking, i.e.
twice that in Eq.(\ref{HawkingTvelocityProbability}). 

From our point of view the existence of Hawking  radiation and the Hawking temperature
depend on the vacuum state. In our case when the horizon is formed by
the moving superfluid liquid, the initial vacuum is the comoving vacuum, i.e. the vacuum
seen in the frame moving with the liquid. The comoving vacuum does not coincide with the
equilibrium vacuum  determined by the Killing vector energy (the vacuum which is seen in
the laboratory frame). That is why the comoving vacuum is unstable, and the initial
stage of the development of instability is the Hawking radiation (see discussion in Ref.
\cite{VolovikBook}).  These two vacua and the quantum tunneling are described in the
Painlev\'e-Gullstrand frame, which is the natural frame for the propagation of
quasiparticles in the moving liquids.

On the contrary, the Schwarzschild coordinates  are irrelevant for this problem: they
are obtained after singular coordinate transformation which is unphysical in
superfluids since it removes the part of the physical space. This is the typical
situation in condensed matter, that in general the singular coordinate transformations
are forbidden. As a result the coordinate frames are distributed into classes: the
transition between the classes is only possible after singular ccordinate
transformations. For the artificial (acoustic) black holes
where the proper frame is the Painlev\'e-Gullstrand frame and the proper initial vacuum
is the comoving vacuum, the Hawking radiation is the real process. The original
Hawking's derivation
\cite{Hawking} has been made in the frame which belongs to the same class as the
Painlev\'e-Gullstrand frame. That is why the radiation from the artificial black hole is
characterized by the same temperature as was derived by Hawking. However, the
Hawking radiation from the astrophysical black hole stiil remains the open question
since the problem of the correct choice of the vacuum state after collapse is not
solved.

Note also that for quantum liquids, in some cases  the Hawking radiation  is suppressed
because of the effect which is similar to the Coulomb blockade in mesoscopic physics:
the radiated quasiparticle disturbs the vacuum state so that the radiation becomes
impossible. For example, the radiated quasiparticle changes the superfluid current and
thus violates the mass conservation law. However the radiation of two quasiparticles is
possible, since it does not violate the conservation law. The simultaneous tunneling of
two quasiparticles is called the co-tunneling. The corresponding tunneling exponent
describing the co-tunneling process is twice the tunneling exponent of a single
quasiparticle. As a result the  Hawking temperature is twice smaller than in
Eq.(\ref{HawkingTvelocityProbability}), i.e the Coulomb blockade effect is opposite to
that suggested in Ref. \cite{Akhmedov}. 

\section{GRAVITY FOR RIPPLONS}
\label{SecGravityRipplons}

In all the condensed matter systems suggested  for simulation of
gravity, even if the horizon can be constructed in principle,  the
estimate for the Hawking temperature is very pessimistic for the
observation of the Hawking radiation  (though the classical analog of the
Hawking effect discussed in the previous section can be observed). That
is why it is instructive to look for the other possible experimental
consequences of the event horizon. The effective gravity for ripplons --
the capillary-gravity waves on the free surface of the liquid or at the
interface between two liquids -- provides us with the new mechanism of
the decay of the black hole. Let us start with ripplons propagating on
the surface of the liquid.

\subsection{Effective Metric on the Surface of Liquid}
\label{SecEffectiveMetric}

The general dispersion relation $\omega({\bf k})$ for ripplons on  the
surface of a  liquid is
\begin{equation}
M(k) (\omega- {\bf k}\cdot {\bf v})^2
=  F+k^2\sigma ~.
\label{GeneralRipplonSpectrum}
\end{equation}
Here  $\sigma$ is the
surface tension; $F=\rho g$
is the gravity force where $\rho$ is mass density of the liquid; and
${\bf v}$ is the velocity of the liquid.  The quantity $M(k)$ is the
$k$-dependent  mass of the liquid which is forced into motion by the
oscillating surface:
\begin{equation}
M(k)={\rho \over k ~{\rm tanh}~kh}~, \label{Mass}
\end{equation}
where $h$ is the thicknesses of the layer of the liquid.

The spectrum (\ref{GeneralRipplonSpectrum}) becomes ``relativistic''  in
the shallow water limit $kh\ll 1$,
$k\ll k_0$:
\begin{equation}
(\omega- {\bf k}\cdot {\bf v})^2
=  c^2k^2 + c^2k^4\left(\frac{1}{k_0^2} -  \frac{1}{3} 
h^2\right)~~,~~c^2=gh~~,~~k_0^2=\rho g/\sigma ~.
\label{RelatSpectrum}
\end{equation}
If the $k^4$ corrections are ignored, the spectrum of ripplons is
described by the effective 2+1 metric \cite{SchutzholdUnruh} 
\begin{equation}
g^{\mu\nu}k_\mu k_\nu=0~~,~~k_\mu=(-\omega, k_x,k_y)~,
\label{EffectiveMetric}
\end{equation} 
with the following elements
\begin{equation}
g^{00}=-1~,~g^{0i}=-v^i~,~g^{ij}=c^2\delta^{ij}-v^iv^j~.
\label{EffectiveMetricContravariant}
\end{equation} 
The interval describing the effective 2+1 space-time in which  ripplons
propagate along geodesics  and the corresponding covariant components of
the effective metric are
\begin{equation}
ds^2=g_{\mu\nu}dx^\mu dx^\nu~~,~~g_{00}=-1
+\frac{v^2}{c^2}~,~g_{0i}=-\frac{v^i}{c^2}~,
~g_{ij}=\frac{1}{c^2}\delta_{ij}~.
\label{EffectiveMetricCovariant}
\end{equation}
 
\subsection{Effective Metric on the Interface Between Two Superfluids}
\label{SecEffectiveMetricInterface}

The spectrum of  ripplons propagating along the interface between two
superfluids:
\begin{equation}
M_1(k) (\omega- {\bf k}\cdot {\bf v}_1)^2
+ M_2(k)(\omega- {\bf k}\cdot {\bf v}_2)^2
=  F+k^2\sigma ~.
\label{GeneralRipplonSpectrumAB}
\end{equation}
Here, as before, $\sigma$ is the surface tension of the interface and $F$
is the force stabilizing the position of the interface. For the interface
between $^3$He-A and $^3$He-B the gravity force is negligible, since
these liquids have almost equal densities: $|\rho_1-\rho_2| \sim
10^{-5}(\rho_1+\rho_2)$. However, these two superfluids have
essentially different magnetic properties, and the corresponding force $F$
which stabilizes the position of the interface is provided by the
gradient of external magnetic field. The effective masses 
$M_1(k)$ and $M_2(k)$ are :
\begin{equation}
M_1(k)=\frac{\rho_{{\rm s}1}} {k  ~{\rm
tanh}~kh_1}~,~M_2(k)=\frac{\rho_{{\rm s}2}}{  k~ {\rm tanh}~kh_2}~,
\label{Masses}
\end{equation}
where $h_1$ and $h_2$ are the thicknesses of the layers of two
superfluids; 
$\rho_{{\rm s}1}$ and $\rho_{{\rm s}2}$ are superfluid densities of the
liquids. Further we assume that the temperature is low enough so that the
normal fraction of each of the two superfluid liquids is small, then
$\rho_{{\rm s}1}\approx \rho_1$ and  $\rho_{{\rm s}2}\approx \rho_2$.

Experiments with the AB interface
\cite{Kelvin-HelmholtzInstabilitySuperfluids} where conducted in the
`deep water' limit   $kh_1\gg 1$ and $kh_2\gg 1$. In the opposite
limit of a thin slab, where $kh_1\ll 1$ and $kh_2\ll 1$, one obtains
\begin{equation}
\alpha_1  (\omega- {\bf k}\cdot {\bf v}_1)^2
+ \alpha_2(\omega- {\bf k}\cdot {\bf v}_2)^2
= c^2k^2  ~,
\label{ThinSlabsSpectrum}
\end{equation}
where
\begin{equation}
\alpha_1  = {h_2\rho_1\over h_2\rho_1+h_1\rho_2}~,~
\alpha_2  ={h_1\rho_2\over h_2\rho_1+h_1\rho_2}~,~c^2= {Fh_1h_2\over
h_2\rho_1+h_1\rho_2}~.
\label{ThinSlabsSpectrumParameters9}
\end{equation}
This can be
rewritten in the Lorentzian form (\ref{EffectiveMetric}) with the
following effective contravariant metric
$g^{\mu\nu}$:
\begin{equation}
g^{00}=-1~,~g^{0i}=-\alpha_1 v_1^i - \alpha_2
v_2^i~,~g^{ij}=c^2\delta^{ij}-\alpha_1 v_1^iv_1^j - \alpha_2 v_2^i v_2^j~.
\label{ThinSlabsRipplonContravMetric}
\end{equation}

\subsection{Interaction with Environment}
\label{SecEffectiveMetricInterface}

The spectra (\ref{GeneralRipplonSpectrum}) and
(\ref{GeneralRipplonSpectrumAB}) are valid for the perfect fluid, where
dissipation due to friction and viscosity is neglected. They must be
modified when the dissipation is added. For the ripplons propagating at
the interface between two superfluids the dissipation leads to a simple
extra term on the right-hand side of Eq.(\ref{GeneralRipplonSpectrumAB})
\cite{VolovikBook,Volovik}:
\begin{equation}
M_1(k) (\omega- {\bf k}\cdot {\bf v}_1)^2
+ M_2(k)(\omega- {\bf k}\cdot {\bf v}_2)^2
=  F+k^2\sigma -   i \Gamma \omega~.
\label{GeneralRipplonSpectrumAB2}
\end{equation}
For the ripplons at the interface between $^3$He-A and $^3$He-B the
friction parameter $\Gamma>0$ depends on temperature and is proportional 
to $T^3$ at low $T$ \cite{KopninInterface}.  The important property of
the added dissipative term is that it introduces the reference  frame of
the environment. The $\omega$-dependence of the dissipative term in Eq.
(\ref{GeneralRipplonSpectrumAB2}), which has no Doppler shift,  implies 
that this spectrum  is written in the frame of the container.

Since the main effect of this term is the introduction of the
distinguished reference frame, the explicit form of this term is not
important. That is why we expect that similar dissipative interaction with
the environment exists for ripplons propagating on the free surface of
the superfluid and even on the surface of normal viscous liquid:
\begin{equation}
M(k) (\omega- {\bf k}\cdot {\bf v})^2
=  F+k^2\sigma -   i \Gamma \omega~.
\label{GeneralRipplonSpectrum2}
\end{equation}
The parameter $\Gamma$ can be considered as phenomenological parameter,
which depends on $\omega$, $k$ and the Reynolds number of the flowing
liquid.

\subsection{Instability in the Ergoregion and Landau Criterion}
\label{Instability}

In superfluids the stationary (time-independent) flow of superfluid
component is frictionless until the critical velocity of flow is reached.
This is the essence of the phenomenon of superfluidity. However, the
notion of critical velocity implies that there is a preferred reference
frame with respect to which the velocity is counted. This is the reference
frame of the environment (the frame of  the container). In equations, this
reference frame is provided by the
$\Gamma$-term, which describes the dissipative interaction with the 
environment. This term also gives rise to the attenuation of ripplons.

When this term is taken into account, from the spectrum 
$\omega(k)$ in Eq. (\ref{GeneralRipplonSpectrum2}) it follows that the
instability to the formation of the surface waves occurs when the
velocity $v$ of the flow with respect to the wall exceeds the critical
velocity $v_L$, at which  the imaginary part ${\rm Im}~\omega(k_c)$
of the energy spectrum of the critical ripplon with momentum $k_c$
crosses zero and becomes positive at $v>v_L$. At that moment the
attenuation of ripplons transforms to amplification; and critical ripplons
start to grow exponentially 
\cite{VolovikBook,Volovik}. 

The critical velocity $v_L$ and the momentum of the critical ripplon $k_c$
do not depend on the friction parameter $\Gamma$. Moreover, at this
velocity the real part ${\rm Re}~\omega({\bf k})=\omega^{\rm com}(k)+{\bf
k}\cdot {\bf v}_{\rm s}$ also crosses zero: it becomes negative at
$v>v_L$ (see Fig. \ref{TwoCritVelocitiesFig} for ripplons at the AB
interface, which we will discuss in Sec.\ref{KHInstability}).  This means
that the threshold velocity coincides with the Landau criterion for the
ripplon nucleation:
\begin{equation}
v_L ={\rm min}_k \frac{\omega^{\rm com}(k)}{k}
~~,~~\omega^{\rm com}(k)=\sqrt{(\rho g +\sigma k^2)/M(k)} ~.
\label{LandauVel}
\end{equation}
The Landau critical velocity is different  in the ``relativistic''
shallow-water and ``non-relativistic'' deep-water regimes:
\begin{eqnarray}
v_L=c~~,~~k_c=0~~,~~ ~~{\rm if}~~ hk_0 < \sqrt{3}~,
\label{CritVelRel}\\
v_L= c\sqrt{2/hk_0}~~,~~k_c=k_0 ~~,~~{\rm if}~~ hk_0 \gg 1~.
\label{CritVelNonRel}
\end{eqnarray}
In both regimes the frequency of the critical ripplon is $\omega(k_c)=0$, 
i.e. the critical ripplon must be stationary in the wall frame.

The region in the liquid, where the flow velocity
$v$ exceeds  $v_L$, represents the ergoregion, since in the container 
frame the energy of the critical ripplon is negative in this region,
$\omega^{\rm com}(k)+{\bf k}\cdot {\bf v} <0$.
For the ``relativistic'' ripplons, the ergoregion is the region where $v$
exceeds  $c$. This can be expressed in terms of the effective metric in
Eq.(\ref{EffectiveMetricCovariant}): in the ergoregion the metric element
$g_{00}$ changes sign and becomes positive. If the flow is perpendicular
to the  ergosurface (the boundary of the ergoregion), then the
ergosurface serves as the event horizon for ripplons. It is the black
hole horizon, if the liquid moves into the ergoregion, and it is the white
hole horizon if the liquid moves from the ergoregion. 
However, the exponential growth of ripplons in the
supercritical region is universal and does not depend on direction of the
flow in the vicinity of the ergosurface. The flow can represent
the white-hole or black-hole horizons; the relativistic ergoregion without
the horizon; and even the ergogerion for the non-relativisic
quasiparticles.  

The instability develops due to the interaction with the fixed
reference frame and occurs in the region where the energy of the critical
fluctuation is negative in this frame. Such kind of instability is also
called the Miles instability.
\cite{SchutzholdUnruh2} In principle, Miles instability may take  place
behind the horizon of the atsronomical black holes  if there exists the
fundamental reference frame related for example with Planck physics
\cite{VolovikBook,SchutzholdUnruh3}. It  may lead to the decay of the
black hole much faster than the decay due to Hawking radiation.

\subsection{Ergoregion Instability vs Kelvin-Helmholtz Instability}
\label{KHInstability}

In the case of the interface between $^3$He-A and $^3$He-B, the
critical velocity of instability towards the growth of critical  ripplon
has been measured in the nonrelativistic  deep-water regime
\cite{Kelvin-HelmholtzInstabilitySuperfluids}, and has been found in a
good agreement with  the theoretical estmate of the Landau velocity
(modified for the case of two liquids \cite{VolovikBook,Volovik}) wihout
any fitting parameter (see Ref. \cite{Krusius} where also the turbulence triggered in
bulk $^3$He-B by the interface instabilty is discussed). The important additional
feature, which takes place in case of the interface between two sliding superfluids, is
the existence of two different critical velocities. 

\begin{figure} [t]
\centerline{\includegraphics[width=80.mm]{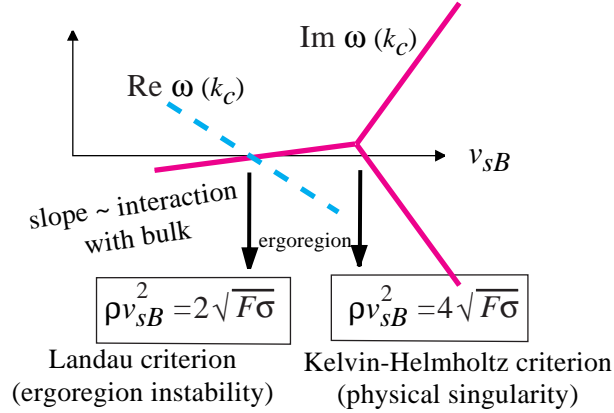}}
%
\caption{Two critical velocities for ripplons living at the
interface between two sliding superfluids -- A and B phases of $^3$He. The lowest
one is the Landau critical velocity, at which the energy of ripplons first becomes
negative.  The region with negative energy is called the ergoregion. If 
the friction parameter $\Gamma$  describing interaction with the
environment is non-zero, the imaginary part of the frequency becomes
positive in the ergoregion,  i.e. attenuation of ripplons transforms
to the amplification in the ergoregion which indicates the instability with
increment  proportional to $\Gamma$. The second velocity marks the more crucial
Kelvin-Helmholtz instability at which the imarginary parts of both signs
emerge. In the `relativistic' shallow-water regime  this velocity
corresponds to physical singularity in metric field (Fig.
\ref{HorizonABbraneFig}). The shown values of two critical velocities
correspond
$T\ll T_c$, deep-water limit, and the experimental condition for the flow
of two superfluids: the A-phase is stationary with respect to the walls of
container, and the B-phase is moving with respect to the container with
the velocity $v_{sB}$. } 
\label{TwoCritVelocitiesFig}
 \end{figure}

One of them is the Landau critical velocity, above which the ergoregion
appears for ripplons. In the typical situation the A-phase is stationary
with respect to the walls of container, while the B-phase is moving with
respect to the container with the velocity
$v_{sB}$. For the deep-water regime   
and in the limit of low temperature $T\ll T_c$, this  critical velocity
is given by: 
\begin{equation}
v_L^2 = \frac{2}{\rho}\sqrt{F\sigma} ~.
\label{LandauVelAB}
\end{equation}
The second critical velocity occurs if one completely neglects the
interaction with the enviroment putting the friction parameter $\Gamma=0$.
In this case the preferred reference frame disappears from the
equations, and thus the physics is determined by the relative
velocity ${\bf v}_1-{\bf v}_2$ of two superfluids. In this case it is well
known that the Kelvin-Helmholtz instability of the interface
occurs when the velocity difference reaches some threshold value. In our
case, when the A-phase is stationary with respect in the container frame,
while the B-phase is moving with respect to the container with the
velocity
$v_{sB}$, the relative velocity of two superfluids is $|{\bf v}_1-{\bf
v}_2|=v_{sB}$ and the Kelvin-Helmholtz criterion becomes
\begin{equation}
v_{KH}^2 = \frac{4}{\rho}\sqrt{F\sigma} ~.
\label{KHVelAB}
\end{equation}
Here we took into account that the densities of two liquids are equal
$\rho_1=\rho_2=\rho$. The difference between two critical velocities in
such an arrangement is
$\sqrt{2}$.

\begin{figure}[t]
\centerline{\includegraphics[width=\linewidth]{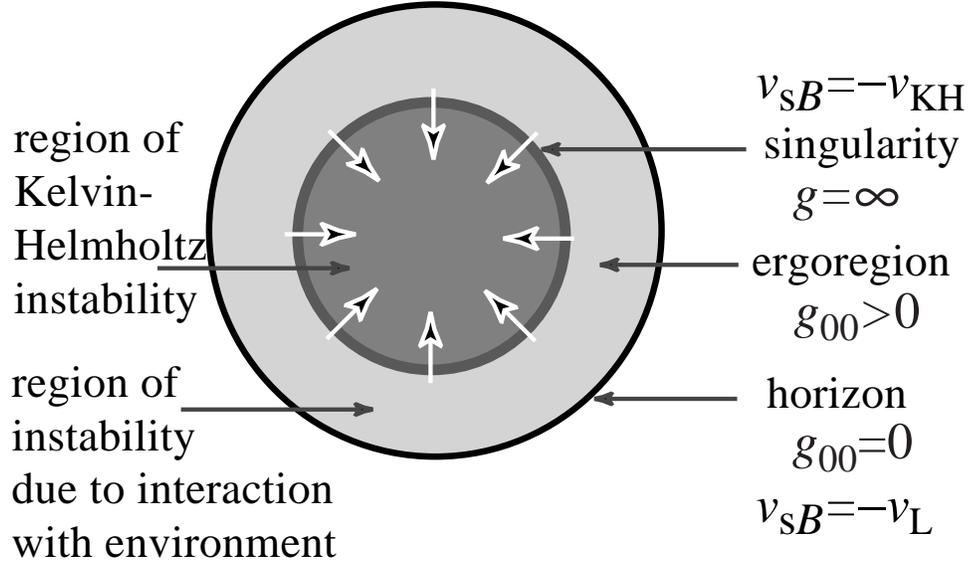}}
\medskip
\caption{Artificial 2+1 black hole for ripplons at the AB interface in the
``relativistic'' shallow-water limit, which can occur for the radial
flow of the B-phase near the interface.  The black hole horizon occurs
when the flow reaches the Landau critical velocity
$|v(r)|=v_L$. At larger critical velocity
$|v(r)|= v_{KH} >v_L$ the Kelvin-Helmholtz instability occurs,  which in
terms of effective metric corresponds to the physical singularity within
the black hole.}
  \label{HorizonABbraneFig}
\end{figure}

The behavior of the real and imaginary parts of the spectrum of the
crtitical ripplon as functions of $v_{sB}$ is shown in Fig.
\ref{TwoCritVelocitiesFig}. When  $\Gamma\neq 0$, at $v_{KH}$ one has 
the smooth crossover, which becomes the true threshold at $\Gamma= 0$.
In the ``relativistic'' shallow-water regime, both thresholds have
physical meaning, especially when the flow is radial (Fig. \ref{HorizonABbraneFig}). The circumference  $v(r)=-v_L=-c/\sqrt{\alpha_1}$ represents the black-hole
horizon. Deep inside the black hole, at
$v(r)=-v_{KH}=-c/\sqrt{\alpha_1\alpha_2}$, the real singularity takes place:
the $g_{rr}$ and $g_{0r}$ components become infinite together with the
determinant $g$ of the  metric $g_{\mu\nu}$.

\section{HYDRAULIC JUMP AS WHITE HOLE}
\label{WH}

 \subsection{Hydraulic Jump}
\label{HydraulicJumpSec}

\begin{figure}[t]
\centerline{\includegraphics[width=110.mm]{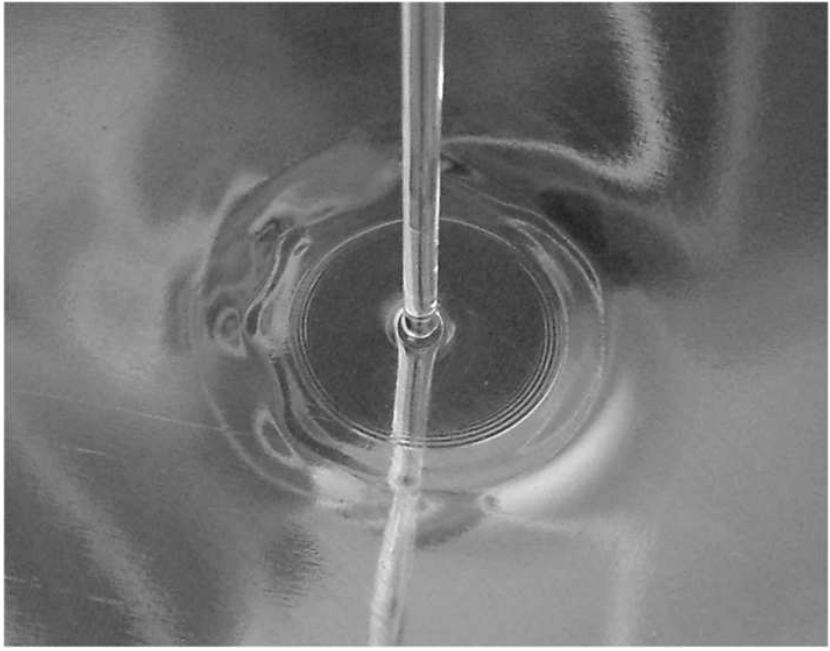}}
%
\caption{Hydraulic jump. The standing waves at the otherwise
smooth surface inside the white hole is the result of the egroregion
instability. Courtesy Piotr Pieranski} 
\label{WHFig}
 \end{figure}

The analog of a white hole horizon for the `relativistic' ripplons on the
surface of a shallow liquid is achieved in the kitchen-bath experiments.
This is the so-called hydraulic jump first discussed by Rayleigh in terms
of the shock  wave
\cite{Rayleigh}. The circular hydraulic jump occurs when the vertical jet
of water falls on a flat horizontal surface (Fig. \ref{WHFig}). The flow
of the liquid at the surface exhibits a ring discontinuity at a certain
distance
$r=R$ from the jet  (observation of the non-circular hydraulic  jumps with
sharp corners has been reported in Ref. \cite{Ellegaard1}). At $r=R$
there is an abrupt increase in the depth $h$ of the liquid (typically by
order of magnitude) and correspondingly a decrease in the radial velocity
of the liquid. The velocity of the liquid in the interior region ($r<R$)
exceeds the speed of `light'  for ripplons
$v>c=\sqrt{hg}$, while outside the hydraulic jump ($r>R$) one has
$v<c=\sqrt{hg}$ (Fig. \ref{HydraulicJumpInstabilityFig}). Since the
velocity flow is radial and outward, the interior  region imitates the
`white-hole' region.  The  interval of the 2+1 dimensional effective
space-time in which the ``relativistic''  ripplons ``live''  is
\begin{equation}
ds^2=-c^2dt^2  + (dr-v(r)dt)^2 +r^2d\phi^2~.
\label{Interval} 
\end{equation}

The surface inside the white hole is smooth, since all the perturbations
are removed by `superluminal' flow of the liquid.
In general relativity the metric  is continuous across the horizon. In
our case there is a real physical singularity at the white-hole horizon
-- the jump in the effective metric (\ref{EffectiveMetricCovariant}). 
However, the discussed mechanism of the Miles instability in the
ergoregion (or behind the horizon) discussed in Sec.
\ref{Instability} towards the growth of the critical ripplon does not
depend on whether the horizon or ergosurface  is smooth or singular. 
According to this instability the critical ripplon must be stationary in
the wall frame. The standing waves inside the white hole are clearly seen
in Fig. \ref{WHFig}. The exponential growth of the critical ripplon is
saturated by the non-linear effects, and then the whole pattern remains
stationary (time-independent but not static).

A similar 3D analog of the black or white hole 
with the physical singularity at the horizon has been
discussed by Vachaspati \cite{Vachaspati}. The role of the
physical horizon is played by the phase boundary between two quantum
vacua -- two superfluid phases of the same liquid, such as $^3$He-A and
$^3$He-B. Two vacua have different `speed of light', say, different
speed of sound, $c_1\neq c_2$.   The analog of black or white horizon
occurs if the superfluid velocity of the flow through the phase boundary
is subsonic in one of the superfluids but supersonic in the other one,
$c_1<v<c_2$. 

\subsection{Hydraulic Jump in Superfluids}
\label{HJ}

\begin{figure} [t]
\centerline{\includegraphics[width=100.mm]{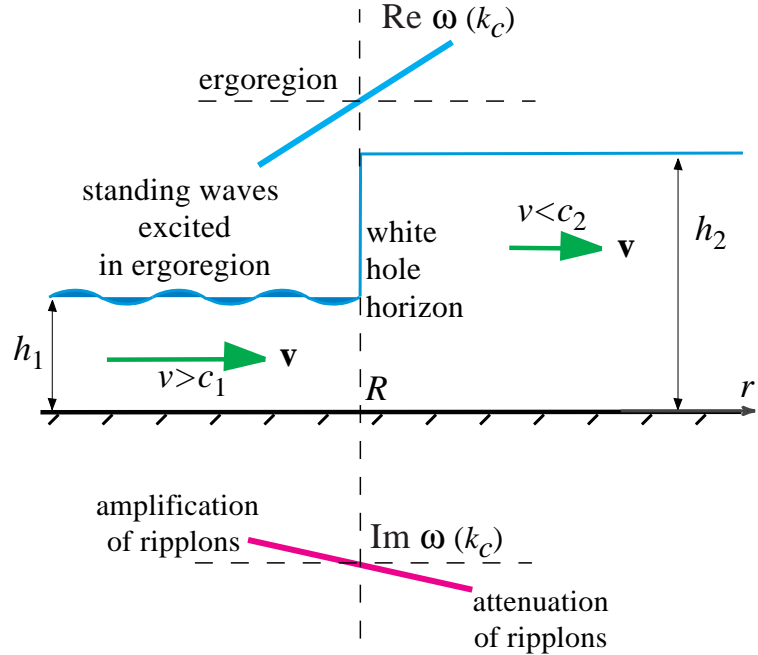}}
%
\caption{Instability in the ergoregion. At the horizon both the energy
and imaginary part of the ripplon spectrum cross zero. In the ergoregion
where the energy is negative, the imaginary part is positive indicating
instability towards the growth of the critical ripplon.} 
\label{HydraulicJumpInstabilityFig}
 \end{figure}

The analogy between the instability of the  surface inside the hydraulic
jump and the instability of the vacuum behind the horizon can be useful
only if the liquid simulates the quantum vacuum. For that, the liquid
must be quantum, and its flow should not exhibit any friction in the
absence of a horizon. That is why the full analogy could occur if one
uses either the flow of quantum liquid with high Reynolds number, or the
superfluid liquid which has no viscosity. Quantum liquids such as
superfluid or normal $^3$He and $^4$He are good candidates.

The first observation of the circular hydraulic  jump in superfluid
liquid (superfluid $^4$He) was reported in Ref. \cite{Pettersen}. The
surface waves generated in the ergoregion (in the region inside the jump
in Fig. \ref{HydraulicJumpInstabilityFig}) were also observed.
As in the normal liquid, this instability is saturated by the non-linear
terms. The same instability observed at
the interface between A-phase and B-phase
 has a different consequence \cite{Kelvin-HelmholtzInstabilitySuperfluids}:
the instability is not saturated and leads to the crucial rearrangement
of the vacuum state: quantized vortices start to penetrate into the 
$^3$He-B side from $^3$He-A \cite{ROPrevirew}. They partially or
fully screen the
$^3$He-B flow and reduce its velocity back below the threshold  for the
ripplon formation. Thus the instability kills the ergoregion.

Under the conditions of experiment   \cite{Pettersen} the hydraulic jump
in superfluid $^4$He is very similar to that in the normal liquid $^4$He.
The position $R$ of the hydraulic jump as a function of temperature  does
not experience discontinuity at the superfluid transition. This suggests 
that quantized vortices are present in the flow providing the mutual
friction between the superfluid and normal components of the liquid. As a
result even below the $\lambda$-point,  the liquid moves as a whole though
with lower viscosity because of the reduced fraction of the normal
component.

To avoid the effect of the normal component  it would be desirable to
reduce the temperature or to conduct similar experiments in a shallow
superfluid $^3$He. 

The advantage of superfluid $^3$He is that,  as distinct from the
superfluid $^4$He, vortices are not easily formed there:  the energy
barrier for vortex nucleation in $^3$He-B is about
$10^6$ times bigger than temperature \cite{Parts}.  In addition, in
superfluid $^3$He  the normal component of the liquid is very viscous
compared to that in superfluid $^4$He. In the normal state the kinematic
viscosity is $\nu\sim 10^{-4}$ cm$^2$/s in liquid $^4$He, and $\nu\sim 1$
cm$^2$/s in liquid $^3$He.    That is why in  many practical arrangements
the normal component in superfluid $^3$He remains at rest with respect to
the reference frame of the wall and thus does not produce any
dissipation  if the flow of the superfluid component is sub-critical. 

One can also exploit thin films of a superfluid liquid, where the normal
component is fixed. The ripplons there represent the so-called third
sound (recent discussion on the third sound propagating in superfluid
$^3$He  films can be found in Ref. \cite{Sauls}).  In 1999 Seamus Davis
suggested to use the third sound in superfluid $^3$He for simulation of
the horizons \cite{Davis}.

In normal liquids it is the viscosity which determines  the position $R$
of the hydraulic jump (see \cite{Bohr}). The open question is what is the
dissipation mechanism which determines the position $R$ of the white-hole
horizon in a superfluid flow with stationary or absent normal component
when its viscosity is effectively switched off.  Since there is no
dissipation of the superfluid flow if its velocity is below $v_L$, one
may expect that  the same mechanism, which is responsible for dissipation
in the presence of the horizon, also determines the position $R$ of the
horizon. If so, the measurement of $R$ as function of parameters of the
system will give the information on various mechanisms of decay of white
hole.  It is also unclear whether  it is possible  to approach the limit
of a  smooth horizon,  without the shock wave of the hydraulic jump; and
whether it is possible to construct the inward flow of the liquid which
would serve as analog of the black hole horizon. 

\section{CONCLUSION}
\label{ConclusionSec}

It appears that the ripplons -- gravity-capillary waves on the surface of
liquids or at the interfaces between superfluids -- are the most
favourable excitations for simulation of the effects related to horizons
and ergoregions. The white-hole horizon for the ``relativistic'' ripplons
at the surface of the shallow liquid is easily simulated using the
kitchen-bath hydraulic jump. The same white-hole horizon is observed in
quantum liquid -- superfluid $^4$He  \cite{Pettersen}. The ergoregion for
the ``non-relativistic'' ripplons is generated in the experiments with two
sliding superfluids \cite{Kelvin-HelmholtzInstabilitySuperfluids}. 

The
common property experienced by the ripplons in all these cases is the
Miles instability inside the ergoregion or horizon. In some cases these
instability is saturated leading to the standing waves inside the horizon
or ergoregion, while in the other cases, such as for ripplons at the
interface between $^3$He-A and  $^3$He-B, the instability leads to the
complete elimination of the ergoregion. Because of the universality of
the Miles instability, one may expect that it could take place inside the
horizon of the astrophysical black holes, if there is a preferred
reference frame which comes from the trans-Planckian physics (see Sec.
32.4 in Ref.\cite{VolovikBook}). If this is the case, the black hole
would evapotate much faster than due to the Hawking radiation. 

I hope, the future experiments in quantum liquids will explore the quantum
limit, where the quantum effects related to horizon and ergoregion become
more pronounced. Experiments with hydraulic jump in superfluid $^4$He
must be extended to the low-temperature region $T\ll T_c$, while the
experiments with ripplons at the interface between $^3$He superfluids must
be extended to the `shallow-water' regime, where ripplons become
`relativistic'.  And, of course, ripplons at the interface
between different superfluid states of ultra-cold Bose and Fermi gases
shall be used.

If the vacuum instability inside
the horizon/ergoregion  is saturated as in experiments with a free surface
\cite{Pettersen}, the other mechanisms will intervene such as the
black-hole laser \cite {BHlaser}, and even the quantum mechanical Hawking
radiation of ripplons. The latter should be enhanced at the sharp
discontinuous horizon of the hydraulic jump and maybe near the sharp
corners of the non-circular (polygonal)  hydraulic jump observed in Ref. 
\cite{Ellegaard1}.

\section*{ACKNOWLEDGMENTS}
I  thank Etienne Rolley and Michael Pettersen  who sent me their
experimental results prior to publication, Marc Rabaud and Tomas Bohr
for discussions and Piotr Pieranski for the beautiful photo of the
hydraulic jump. This work is supported in part by the Russian Ministry of
Education and Science, through the Leading Scientific School grant
$\#$1157.2006.2, and by the European Science Foundation  COSLAB Program.

\end{document}